\documentclass[sigconf,authorversion,nonacm]{acmart}
\usepackage{color}

\usepackage{float}

\usepackage{booktabs}
\usepackage{multirow}

\usepackage[toc,page]{appendix} 
\usepackage{tabularx}
\usepackage{url}
\usepackage{enumitem}
\setlist[itemize]{leftmargin=*}

\usepackage[ruled]{algorithm2e}

\AtBeginDocument{%
  \providecommand\BibTeX{{%
    \normalfont B\kern-0.5em{\scshape i\kern-0.25em b}\kern-0.8em\TeX}}}

\begin{document}



\title{Beyond Recommender: An Exploratory Study of the Effects of Different AI Roles in AI-Assisted Decision Making}

\author{Shuai Ma}
\orcid{0000-0002-7658-292X}
\affiliation{
  \institution{The Hong Kong University of Science and Technology}
  \city{Hong Kong}
  \country{China}
}
\email{shuai.ma@connect.ust.hk}

\author{Chenyi Zhang}
\affiliation{
  \institution{ShanghaiTech University}
  \city{Shanghai}
  \country{China}
}
\email{zhangchy3@shanghaitech.edu.cn}

\author{Xinru Wang}
\affiliation{
  \institution{Purdue University}
  \city{West Lafayette}
  \state{Indiana}
  \country{USA}
}
\email{xinruw@purdue.edu}

\author{Xiaojuan Ma}
\affiliation{
  \institution{The Hong Kong University of Science and Technology}
  \city{Hong Kong}
  \country{China}
}
\email{mxj@cse.ust.hk}

\author{Ming Yin}
\affiliation{
  \institution{Purdue University}
  \city{West Lafayette}
  \state{Indiana}
  \country{USA}
}
\email{mingyin@purdue.edu}

\renewcommand{\shortauthors}{Shuai Ma, et al.}

\begin{abstract}

Artificial Intelligence (AI) is increasingly employed in various decision-making tasks, typically as a Recommender, providing recommendations that the AI deems correct. However, recent studies suggest this may diminish human analytical thinking and lead to humans’ inappropriate reliance on AI, impairing the synergy in human-AI teams. In contrast, human advisors in group decision-making perform various roles, such as analyzing alternative options or criticizing decision-makers to encourage their critical thinking. This diversity of roles has not yet been empirically explored in AI assistance. In this paper, we examine three AI roles: Recommender, Analyzer, and Devil's Advocate, and evaluate their effects across two AI performance levels. Our results show each role's distinct strengths and limitations in task performance, reliance appropriateness, and user experience. Notably, the Recommender role is not always the most effective, especially if the AI performance level is low, the Analyzer role may be preferable. These insights offer valuable implications for designing AI assistants with adaptive functional roles according to different situations.
\end{abstract}

\ccsdesc[500]{Human-centered computing~Empirical studies in HCI}


\keywords{AI-Assisted Decision-making, Human-AI Collaboration, Appropriate Reliance}

\maketitle

\section{Introduction}
With remarkable technological advancements, AI has been increasingly used to support people in making decisions in various domains, ranging from low-stakes recommendations around entertainment \cite{wang2014improving, ma2019smarteye} to high-stakes decision tasks such as criminal justice \cite{dodge2019explaining, dressel2018accuracy}, admissions and employment \cite{cheng2019explaining, zhang2023deliberating}, financial investment \cite{green2019principles}, medical diagnosis \cite{cai2019hello, lee2021human}, etc. Due to AI's imperfect accuracy, and the concerns of safety, ethics, and accountability \cite{cai2019hello, lee2021human, binns2018s}, the \emph{AI-assisted decision-making} paradigm is widely adopted in real-world applications \cite{buccinca2021trust, zhang2020effect, wang2021explanations, bansal2021does}. In this paradigm, AI performs an assistive role by providing a recommendation, while human decision-makers can choose to accept or reject AI's suggestion in their final decision \cite{lai2021towards}.

However, empirical research reveals several limitations within the existing AI-assisted decision-making framework, wherein AI acts primarily as a recommender. One notable issue is that individuals, when passively receiving AI suggestions, seldom engage in analytical thinking \cite{buccinca2020proxy, rastogi2020deciding, bertrand2022cognitive}. Furthermore, people frequently inappropriately rely on the AI's recommendations (such as over-reliance and under-reliance) \cite{buccinca2021trust, ma2023should, wang2021explanations, ma2022glancee} and the mere provision of AI explanations can, paradoxically, exacerbate overreliance \cite{bansal2021does, poursabzi2021manipulating}.

In comparison, in human-human decision-making, beyond recommenders, human advisors sometimes play other types of roles, such as helping the decision-makers analyze the pros and cons of different alternatives instead of directly giving recommendations, or critically challenging decision-makers' initial views \cite{schwenk1984devil, schweiger1989experiential}. Empirical studies show that those roles sometimes help decision-makers achieve better decision-making performance than just giving recommendations \cite{schwenk1984devil, schweiger1986group, cosier1978effects}. Despite the potential of different advisor roles, existing research on AI-assisted decision-making mainly allows AI to play the role of recommender and little is known about the effects of AI playing other roles in AI-assisted decision-making. Specifically, we pose the following research questions.
\begin{itemize}
    \item \textbf{RQ1}: How will different AI assistant roles affect humans' task performance?
    \item \textbf{RQ2}: How will different AI assistant roles affect humans' reliance and the appropriateness of reliance?
    \item \textbf{RQ3}: How will different AI assistant roles affect humans' user experience in the decision-making process?
\end{itemize} 
Additionally, since the advisors' expertise can influence the quality of the support information provided and the decision-makers' receptiveness to varying types of suggestions \cite{dalal2010types, bonaccio2006advice}, we incorporate the factor of AI performance into exploration.


To investigate these research questions, we conducted a user study with participants recruited from Prolific. They were tasked with classifying short news paragraphs into one of three categories: business, entertainment, or politics. We tested three distinct AI roles: (1) Recommender, which provides the AI's recommendation and explanations; (2) Analyzer, which aids in analyzing each option without specific recommendations, offering evidence both \emph{for} and \emph{against} each alternative; (3) Devil's Advocate, where the AI presents counter-arguments to the participant's initial choice, even if it aligns with the AI's own judgment. Additionally, we implemented two levels of AI performance: low (65\% accuracy) and high (90\% accuracy). Therefore, there are six conditions (three AI roles * two levels of AI performance) in total. The experimental results revealed that when AI performance is high, the Recommender role tends to lead to better task performance, more appropriate reliance, and better user experience. However, when AI performance is low, the Analyzer role is more effective in contributing to higher task performance, more appropriate reliance, and better user experience. These findings reveal that there may not be a one-size-fits-all AI role and the choice of AI role should be determined by other factors. Finally, we discussed our findings, the implications for designers, and potential directions for future research.

\section{Related Work}

Artificial Intelligence (AI) is increasingly used in decision-making across various domains \cite{dastin2018amazon, dilsizian2014artificial, khandani2010consumer, wang2014improving, yang2018insurance}. However, AI's real-world applications are not infallible, lacking 100\% accuracy \cite{ma2022modeling, lai2021towards}. This is especially concerning in high-stakes domains like clinical diagnosis and criminal justice, leading to ethical and legal complexities \cite{cai2019hello, lee2021human, binns2018s}. To address this, the prevalent paradigm of AI-assisted decision-making has emerged, drawing substantial attention in the Human-Computer Interaction (HCI) and AI communities \cite{buccinca2021trust, zhang2020effect, wang2021explanations, bansal2021does}. In this paradigm, AI takes on a supportive role, offering recommendations (as well as other information, such as explanations \cite{bansal2021does, zhang2020effect}) for human decision-makers to accept or reject in their final decisions \cite{lai2021towards}.

Empirical research indicates that, in the existing AI-assisted decision-making paradigm, humans often rely on heuristics rather than analytical thinking, making it difficult to appropriately rely on AI's suggestions \cite{buccinca2020proxy, buccinca2021trust}. This contrasts with human-human decision-making, where human advisors can adopt various roles beyond merely recommending. They may assist in weighing the pros and cons of different options, suggesting alternatives, or actively challenging the initial opinions of decision-makers \cite{schwenk1984devil, schweiger1989experiential}. Studies have demonstrated that these roles (such as dialectical inquiry, devil's advocate \cite{schwenk1984devil}) can sometimes enhance decision-making effectiveness more than just providing recommendations \cite{schweiger1986group, cosier1978effects}. However, current research on AI-assisted decision-making predominantly positions AI in the recommender role (suggesting what AI thinks correct), with limited understanding of the impact when AI assumes different advisory roles in the decision-making process.

In human decision-making processes, human advisors often act in three roles: Recommender, Analyzer, and Devil's Advocate. These three roles are the most commonly assumed by human assistants in group or organizational decision processes \cite{schwenk1984devil, schweiger1989experiential}. Moreover, human advisers sometimes take on additional roles. For example, acting as a (1) ``facilitator'' to help decision-makers form or clarify the problem \cite{valacich1995structuring} or teaching decision-makers how to make a decision (for instance, suggesting that making a list of pros and cons could be useful) \cite{schweiger1989experiential}, (2) ``affirmation provider'' because sometimes, decision-makers seek not new options but confirmation that ``you are right'' \cite{sniezek1995cueing}, (3) ``social or emotional supporter'' to give psychological support (e.g., a human advisor may say ``I understand this is really tough for you as choosing a job is a very stressful decision'') \cite{arkes1991organizational, dalal2010types}, and (4) ``information collector'' to provide information unknown to decision-makers or organizing information for them \cite{schweiger1989experiential}. However, since the primary aim of AI-assisted decision-making is to enhance task performance, merely agreeing with the decision-maker or providing emotional support may not be suitable roles for AI to fulfill. Furthermore, this paper focuses on decision-making scenarios based on current task information and does not involve external knowledge, thus eliminating the need for AI to help gather additional information. Therefore, in this paper, we decided to choose the roles of Recommender, Analyzer, and Devil's Advocate for AI to play and investigate the effects on human decision-making.

In recent years, researchers in AI-assisted decision-making and XAI have envisioned having AI play other roles. For example, in the discussion part of Bansal et al.'s work \cite{bansal2021does}, they propose to make AI act as the Devil's Advocate. And in Miller's paper \cite{miller2023explainable}, a new concept called ``hypothesis-driven'' decision support is proposed where the AI will not give suggestions but help users analyze their proposed hypotheses. However, there is a lack of empirical investigation of the actual effects of different AI roles in AI-assisted decision-making. In this paper, we focus on exploring the effects of different AI roles on task performance, the appropriateness of human reliance, and user experience. We also investigate the interaction effects of some key factors, such as AI performance and human-independent performance. We also conduct exploratory post-hoc analysis based on the characteristics of humans.
\section{User Study}
The goal of our study is to empirically understand the effects of different AI roles and the interaction effects between AI roles and AI performance. We begin our study when the task is easy for human decision-makers.

\subsection{Task, AI Model, and AI Explanation}
We selected \emph{news category classification} \cite{misra2022news, misra2021sculpting} as our testbed because text classification has been widely used in several previous studies on human-AI decision-making \cite{lai2019human, lai2020chicago, bansal2021does} and it requires little domain expertise and is thus amenable to crowdsourcing. Specifically, given a piece of short news, participants were asked to classify the news into one of three categories: business, entertainment, and politics.

After processing the textual data (such as removing the stopwords), we trained two AI models to generate AI assistance with two levels of performance. One used a logistic regression model with NLTK tokenizer to train the low-performance AI. The other used a BERT model (bert-base-uncased) to train the high-performance AI.

We used LIME to generate explanations for each model. For each of the three news categories, LIME identified the top 10 important words in a news paragraph and assigned an importance score to each word (a positive score means a positive influence on that category while a negative score means a negative influence). We leveraged words with positive importance scores as AI's collected evidence for a prediction and words with negative importance scores as AI's collected evidence against a prediction.

With pilot studies, we carefully selected 20 news with uniform distributed labels (7 business, 7 entertainment, and 6 politics). Participants' average accuracy on these task instances is around 80\%. The high-performance AI's accuracy is 90\% while the low-performance AI's accuracy is 65\%.

\subsection{Conditions}
We adopt a two-factor between-subjects study design in which each participant sees one of \textbf{three} AI role conditions in one of the \textbf{two} AI performance conditions. So our experiment contains six conditions in total (3*2).

\begin{figure*}[htbp]
	\centering 
	\includegraphics[width=\linewidth]{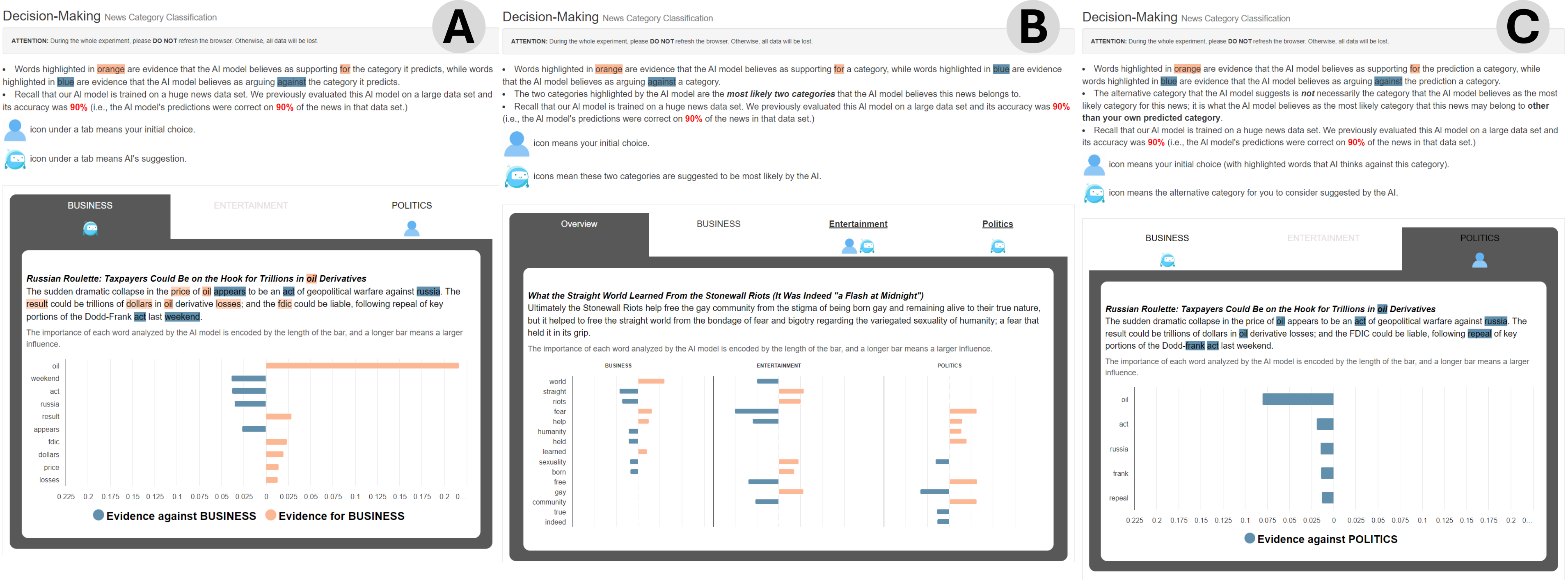}
	\caption{The interfaces of three AI roles. (a) Recommender: In this interface, the AI directly presents a recommendation along with its explanation. (b) Analyzer: The AI assists in evaluating the pros and cons of each option. By default, it displays an overview of all options and highlights two highly probable choices to help users refine their selection. Users can explore detailed AI analyses for any specific option by clicking on its corresponding tab. (c) Devil's Advocate: This interface features the AI providing arguments against the user's current selection, irrespective of whether it aligns with the AI's own prediction. Additionally, the AI suggests an alternative option, which represents the most viable alternative to the user's initial choice.}
	\label{fig:interface}
        \Description{}
\end{figure*}

\noindent\textbf{Factor 1: AI Role}. We compare three types of assistance the AI can provide. Note that the implementation of \textbf{Analyzer} and \textbf{Devil's Advocate} is not fixed, and we do not want to say that the current implementation is the best. We encourage interested researchers to explore more interesting interface designs for these AI roles.
\begin{itemize}
    \item \textbf{Recommender}: The AI model provides a recommendation (which category the AI thinks with the highest probability) and its related explanations (words highlighted for and against this recommendation) (Figure \ref{fig:interface} (a)).
    \item \textbf{Analyzer}: The AI model provides evidence for (words highlighted with positive importance score) and again (words highlighted with negative importance score) each category. To help users narrow the exploration space, the AI model also highlights two options that the AI thinks possible (Figure \ref{fig:interface} (b)).
    \item \textbf{Devil's Advocate}: The AI model provides evidence against what the user predicts (by only highlighting words with negative importance scores). And the AI model marks one option that needs to be explored other than what the user has predicted (Figure \ref{fig:interface} (c)). Note that the AI will always criticize the human - even if the AI agrees with the users' prediction, the AI still provides negative evidence against the user's selected option.    
\end{itemize}
\textbf{Factor 2: AI performance}. The AI performance condition manipulates the accuracy of AI's predictions (although AI predictions are not directly presented to users in some AI role conditions, the quality of AI gathered evidence is affected by AI's accuracy). We note that the definition of performance level can be affected by many factors. For example, in some easy-to-recognize tasks, 90\% accuracy is still deemed as low performance, while in some difficult tasks, even 65\% accuracy can be a reasonable performance. In this experiment, we define AI performance as high or low based on the relative strength between humans and AI. Since from a pilot study, we found that, in this task, humans' accuracy is around 80\%, we artificially set 90\% accuracy as high performance and 65\% accuracy as low performance.
\begin{itemize}
    \item \textbf{High-Performance AI}: The AI model's accuracy on the selected 20 task instances is 90\%.
    \item \textbf{Low-Performance AI}: The AI model's accuracy on the selected 20 task instances is 65\%.   
\end{itemize}
For each AI performance condition, on the study interfaces, we explicitly informed participants of AI's performance - ``The Al model is trained on a huge news data set. We previously evaluated this Al model on a large data set and its accuracy was 90\% (65\%) (i.e., the Al model's predictions were correct on 90\% (65\%) of the news in that data set.)'' In this way, we can avoid any confounding factors that arise from the inconsistency between the stated AI accuracy and the actual AI accuracy.

\subsection{Procedure}
We conducted a between-subjects study. After obtaining consent, participants completed a questionnaire gathering their demographic data, AI expertise, and other characteristics. Participants then went to an interactive tutorial, practiced with one example task, and familiarized themselves with how the AI offers assistance based on the assigned experimental condition. Following the tutorial, qualification questions ensured understanding, with only those answering all questions correctly proceeding to the main task. The main task involved 20 task instances. For each task instance, participants were required to first make their own predictions, then receive AI's assistance, and finally make the final predictions. In the exit survey, we collected participants' user experience.
\subsection{Participants}
After obtaining institutional IRB approval, we recruited participants from Prolific\footnote{www.prolific.co\label{prolific}}. Specific criteria we set for the participants include: (1) a minimum 99\% approval rate for previous submissions; (2) English as their first language; (3) at least 1000 approved previous submissions; and (4) using a desktop computer for the experiment. The study followed a between-subjects design with no repeated participation. After filtering out inattentive participants, we obtained 140 valid responses (more than 20 participants in each condition). Among the participants, there were 75 self-reported males, 63 females, and 2 non-binary individuals. Age distribution included 19 participants aged 18-29, 44 aged 30-39, 42 aged 40-49, 22 aged 50-59, and 13 aged over 59. Participants also self-rated their AI expertise: 17 had no knowledge, 68 knew basic AI concepts, 44 had experience with AI algorithms, and 11 were experts in AI. As an incentive for quality work, participants received a \$1 bonus for an overall accuracy exceeding 95\%. The study took approximately 20 minutes, with participants earning an average wage of \$12 per hour.

\subsection{Measurements and Analysis Methods}
Based on the proposed three research questions, we measured participants' task performance, reliance and the appropriateness of reliance, and user experience.

\noindent\textbf{Task Performance} We calculated participants' initial accuracy before seeing AI's assistance and final accuracy after seeing AI's assistance \cite{ma2023should}.

\noindent\textbf{Reliance and Its Appropriateness} We measure the reliance of participants on the AI system via two metrics \cite{zhang2020effect, he2023knowing}:
\begin{footnotesize}
$$\textnormal{\textbf{Agreement Fraction}} = \frac{\textnormal{Number of final decisions same as the AI suggestion}}{\textnormal{Total number of decisions}},$$

$$\textnormal{\textbf{Switch Fraction}} = \frac{\textnormal{Number of decisions user switched to agree with the AI model}}{\textnormal{Total number of decisions with initial disagreement}},$$
\end{footnotesize}

Based on literature \cite{wang2021explanations, bansal2021does}, we measure the appropriateness of participants' reliance by:
\begin{footnotesize}
$$\textnormal{\textbf{Over-Reliance}} = \frac{\textnormal{Number of incorrect human final decisions with incorrect AI advice}}{\textnormal{Total number of incorrect AI advice}},$$

$$\textnormal{\textbf{Under-Reliance}} = \frac{\textnormal{Number of incorrect human final decisions with correct AI advice}}{\textnormal{Total number of correct AI advice}},$$
\end{footnotesize}

\noindent\textbf{User Experience} We measured participants' \emph{perceived autonomy} \cite{dalal2010types}, \emph{mental demand} \cite{hart2006nasa, ma2023should, ma2022glancee}, \emph{perceived complexity} \cite{buccinca2021trust}, \emph{engagement} \cite{boulton2012predicting}, \emph{future use} \cite{ma2023should, ma2022modeling}, \emph{satisfaction} \cite{buccinca2021trust, ghai2021explainable}, \emph{perceived helpfulness} \cite{laugwitz2008construction, cai2019human, buccinca2020proxy}, \emph{trust} \cite{yin2019understanding}, \emph{self-efficacy} \cite{hemmer2023human} via 7-point Likert scales. The detailed questions are shown in Table \ref{tab:measurement}.
For the data analysis, since the collected user data did not pass the normality test, we used Kruskal-Wallis to compare three AI roles and post-hoc pair-wise comparison with Bonferroni correction.

\renewcommand{\arraystretch}{1.5}
\begin{table*}[htp]  

\centering  
\fontsize{8}{8}\selectfont  

\caption{Questionnaire used in our user study for assessing participants' user experience. All the selected questions are selected and adapted based on verified scales.}\label{tab:measurement}

\begin{tabular}{m{2cm}<{\centering}m{12cm}}
\toprule
\textbf{Aspect}&\textbf{Question}\\ \hline
Autonomy&How much did your interaction with the AI model in these tasks help you maintain the freedom (i.e., autonomy) to make the news category predictions in the way you see fit? (1: Not at all, 7: Extremely)\\
\cline{1-2}
Mental Demand&How mentally demanding was your interaction with the AI model? (1: Very low, 7: Very high)\\
\cline{1-2}
Complexity&The decision-making process with this kind of interface to interact with AI models is complex. (1: Strongly disagree, 7: Strongly agree)\\
\cline{1-2}
Engagement&I am engaged in the decision-making task with the AI model. (1: Strongly disagree, 7: Strongly agree)\\
\cline{1-2}
Future Use&I would like to continue to work with the AI model. (1: Strongly disagree, 7: Strongly agree)\\
\cline{1-2}
Satisfaction&I am satisfied with the AI model's assistance. (1: Strongly disagree, 7: Strongly agree)\\
\cline{1-2}
Helpfulness&I think the AI model’s assistance is helpful/useful for me to make good decisions. (1: Strongly disagree, 7: Strongly agree)\\
\cline{1-2}
\multirow{4}*{\shortstack{Trust}}&I can trust the AI model. (1: Strongly disagree, 7: Strongly agree)\\
\cline{2-2}
&The AI model can be trusted to provide reliable decision support. (1: Strongly disagree, 7: Strongly agree)\\
\cline{2-2}
&I trust the AI model to keep my best interests in mind. (1: Strongly disagree, 7: Strongly agree)\\
\cline{2-2}
&In my opinion, the AI model is trustworthy. (1: Strongly disagree, 7: Strongly agree)\\
\cline{1-2}
\multirow{5}*{\shortstack{Efficacy}}&I am confident about my ability to complete news category classification tasks. (1: Strongly disagree, 7: Strongly agree)\\
\cline{2-2}
&I have mastered the skills necessary for completing news category classification tasks. (1: Strongly disagree, 7: Strongly agree)\\
\cline{2-2}
&I am self-assured about my capabilities to perform news category classification tasks. (1: Strongly disagree, 7: Strongly agree)\\
\bottomrule
\end{tabular}
\end{table*}
\section{Results}
\subsection{Effects of AI roles on task performance}

\begin{figure*}[htbp]
	\centering 
	\includegraphics[width=0.9\linewidth]{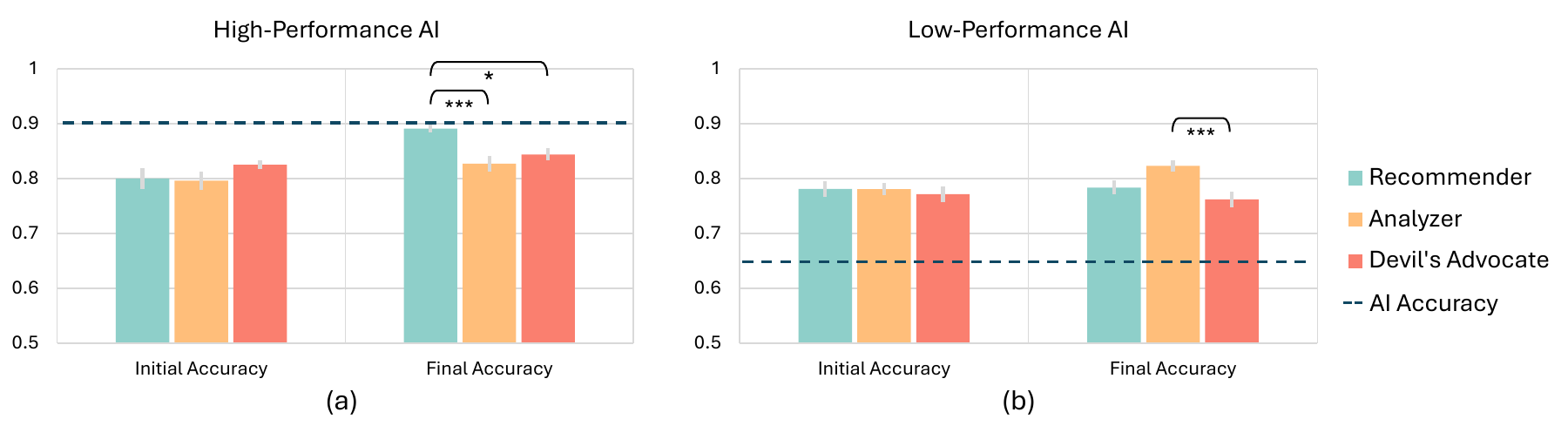}
	\caption{The task accuracy in three AI role conditions. (a) When AI performance is high. (b) When AI performance is low. Error bars represent standard error. (+: $p$<0.1, *: $p$<0.05, **: $p$<0.01, ***: $p$<0.001)}
	\label{fig:accuracy}
        \Description{}
\end{figure*}

Figure \ref{fig:accuracy}(a) demonstrates that with high-performance AI, the Recommender role leads to higher final accuracy compared to both the Analyzer (p<.001) and the Devil's Advocate (p<.05). Conversely, at low AI performance levels, the Analyzer tends to yield higher final accuracy. However, this difference is statistically significant only when compared with the Devil’s Advocate (p<.001). Overall, these findings suggest that high-accuracy AI is best suited for the Recommender role. In cases of suboptimal AI accuracy, it is less appropriate for AI to directly provide recommendations. Instead, it proves more effective in assisting users to analyze the advantages and disadvantages of various decision options.

\begin{figure*}[htbp]
	\centering 
	\includegraphics[width=\linewidth]{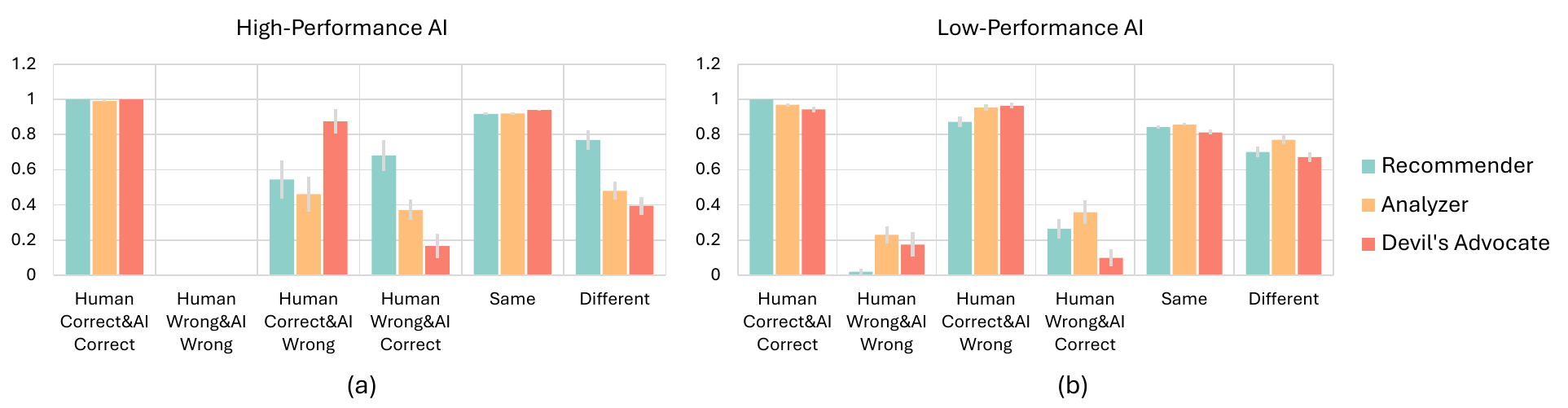}
	\caption{A detailed analysis of participants' final decision accuracy based on the correctness of human initial prediction and AI recommendation. (a) When AI performance is high. (b) When AI performance is low. Error bars represent standard error. (Since categorizing participants' prediction data into different situations reduces the sample size, we did not perform statistical analysis.)}
	\label{fig:detailedanalysis}
        \Description{}
\end{figure*}

We also conducted a detailed analysis of task performance based on the correctness of human initial prediction and AI recommendation. Results in Figure \ref{fig:detailedanalysis} reveal that with high AI accuracy, the Devil's Advocate approach improves human decisions when initial predictions are correct but AI advice is incorrect. Interestingly, under low AI accuracy, both Analyzer and Devil's Advocate surpass the Recommender by aiding better decisions in cases where both human prediction and AI recommendation are wrong, or the human is right but AI is wrong. This indicates that Analyzer and Devil's Advocate remain beneficial for decision-making even when AI recommendations are inaccurate.

\begin{figure*}[htbp]
	\centering 
	\includegraphics[width=\linewidth]{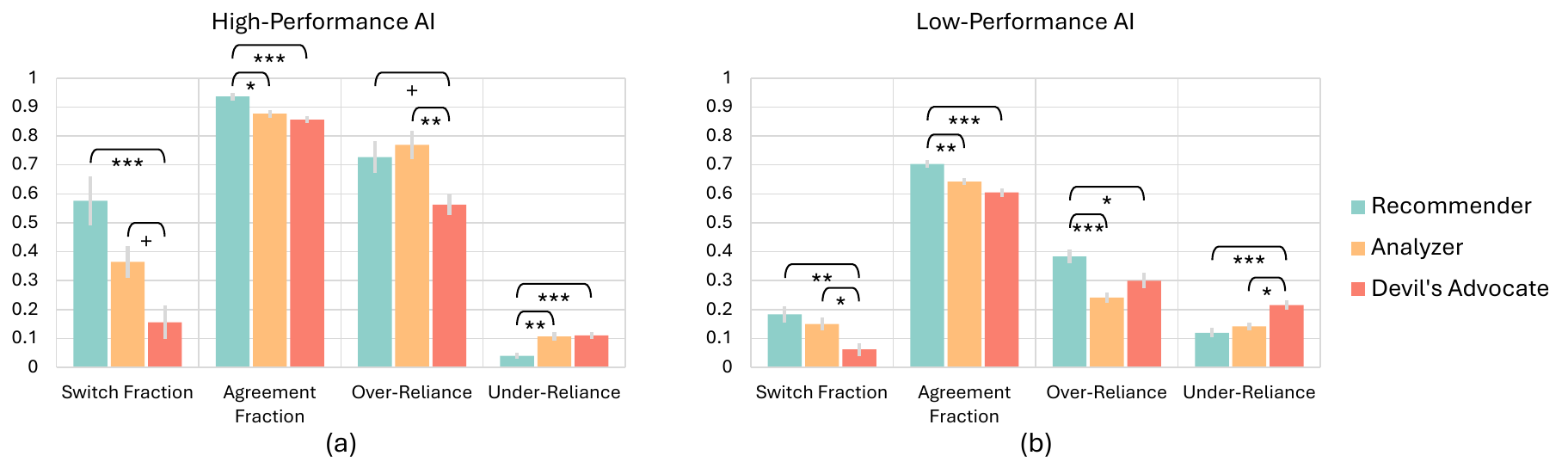}
	\caption{Participants' reliance and the appropriateness of their reliance on AI. (a) When AI performance is high. (b) When AI performance is low. Error bars represent standard error. (+: $p$<0.1, *: $p$<0.05, **: $p$<0.01, ***: $p$<0.001)}
	\label{fig:reliance}
        \Description{}
\end{figure*}

\subsection{Effects on Reliance and the Appropriateness of Reliance}
Figure \ref{fig:reliance} shows that the agreement and switch fractions are highest when AI plays a recommender role, irrespective of AI performance. This outcome is attributed to the Recommender AI's direct presentation of recommendations, in contrast to the implicit decision support provided by the Analyzer and Devil's Advocate. As depicted in Figure \ref{fig:reliance} (a), high AI performance results in lower over-reliance with the Devil's Advocate compared to Analyzer (p<.01). This may be because the Devil's Advocate raises challenges, particularly when AI predictions are incorrect yet align with human predictions, thereby reducing the likelihood of human error. Conversely, Figure \ref{fig:reliance} (b) reveals that with low AI performance, over-reliance is less for the Analyzer and Devil's Advocate than for the Recommender (p<.001 and p<.05 respectively). This is probabaly because, in these roles, AI does not directly present incorrect predictions and participant's lower overall trust in AI (due to the AI's low performance) mitigates the impact of AI misjudgments. Regarding under-reliance, Figure \ref{fig:reliance} (a) shows that high AI performance leads to lower under-reliance with the Recommender role compared to the Analyzer (p<.01) and Devil's Advocate (p<.001). This is due to the Recommender's direct provision of correct predictions, facilitating the acceptance of AI's accurate judgments. Figure \ref{fig:reliance} (b) indicates that low AI performance causes more under-reliance with the Devil's Advocate role, likely because this role prompts people to question even correct AI predictions that align with their views.

In summary, the Devil's Advocate role consistently reduces over-reliance, albeit increasing under-reliance, regardless of AI performance. The Recommender role decreases under-reliance but increases over-reliance. Notably, the Analyzer role diminishes both over-reliance and under-reliance, when AI performance is low. This suggests that the Analyzer role may be more beneficial when AI performance is suboptimal, fostering a more appropriate reliance on AI.

\subsection{Effects on User Experience}

\begin{figure*}[htbp]
	\centering 
	\includegraphics[width=\linewidth]{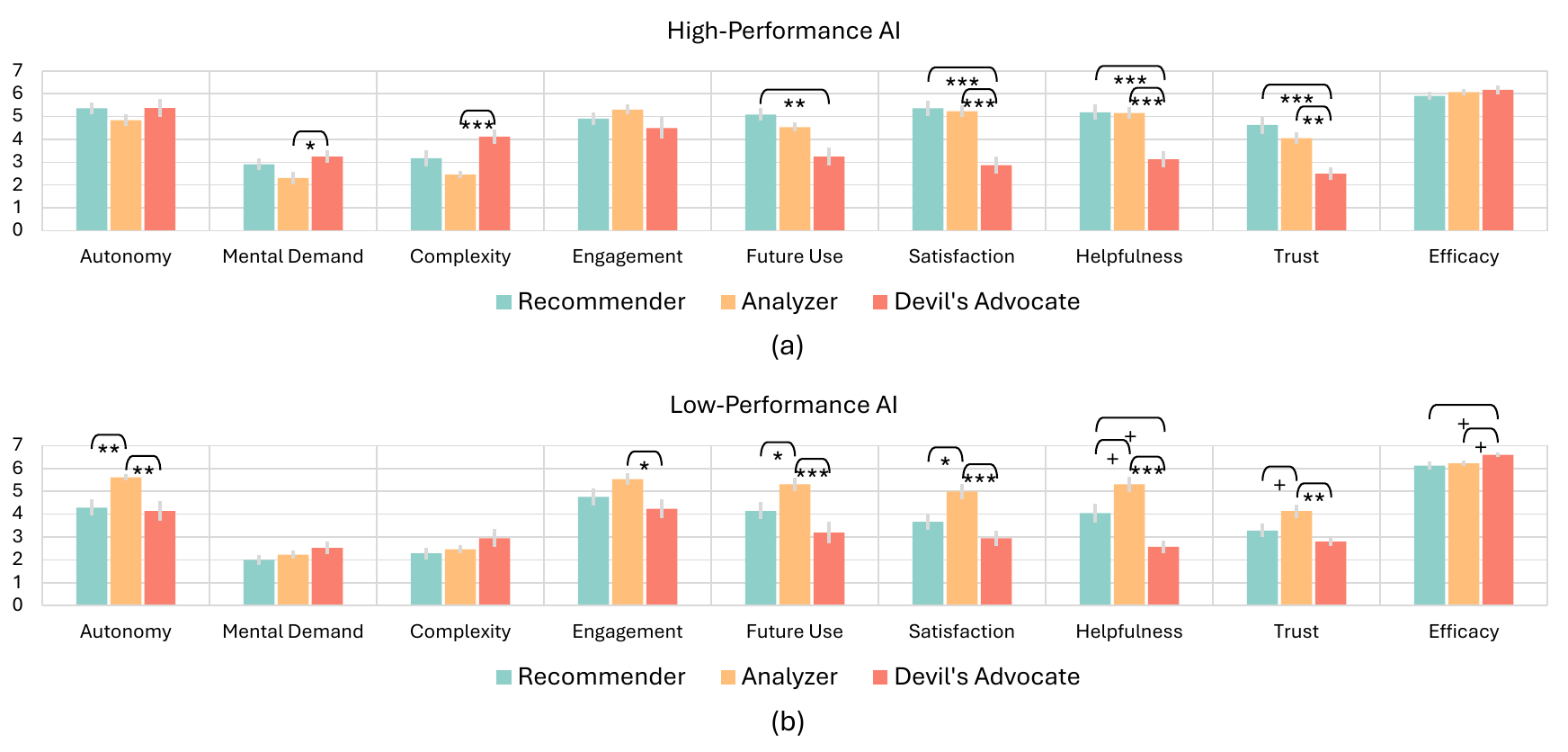}
	\caption{Participants' user experience. (a) When AI performance is high. (b) When AI performance is low. Error bars represent standard error. (+: $p$<0.1, *: $p$<0.05, **: $p$<0.01, ***: $p$<0.001)}
	\label{fig:experience}
        \Description{}
\end{figure*}

Figure \ref{fig:experience} shows the effects of different AI roles on participants' user experience. Results indicate that when AI performance is low, the Analyzer role contributes the most to participants' perceived autonomy. This likely stems from people's reluctance to engage with low-performance AI's suggestions or critiques. Instead, users seem to feel more control when AI just provides analysis. When AI performance is high, the Devil's Advocate role imposes participants' higher mental demands and perceived complexity of the system than Analyzer. This might be due to users' willingness to invest cognitive resources in considering the AI's challenges but they had to take AI's challenges into consideration due to AI's high performance. Moreover, we found that with low AI performance, users engage more with the Analyzer's assistance than with Devil's Advocate. When the AI's accuracy is low, participants show a preference for continuing to work with the Analyzer. In comparison, there is a consistent reluctance to use the Devil's Advocate regardless of AI performance level, possibly because users prefer affirmation over being questioned [ref]. Additionally, when AI performance is high, users exhibit greater satisfaction, helpfulness, and trust in both the Recommender and Analyzer roles. However, when AI performance is low, these positive responses are more pronounced with the Analyzer role. 

In summary, a high-performing AI enhances user experience with both the Recommender and Analyzer roles. In contrast, a lower-performing AI leads to a more favorable user experience when the Analyzer role is employed.

\section{Discussion}
\textbf{Stepping out of XAI-based AI-Assisted Decision-making}. There is currently a growing body of work proposing that AI should provide advice in ways other than just giving suggestions and explanations. For instance, Danry et al. \cite{danry2023don} introduced AI-framed Questioning to enhance critical thinking and human discernment of flawed statements while Miller advocates for AI to ``evaluate'' people's hypotheses instead of giving recommendations \cite{miller2023explainable}. Our research aligns with these innovative approaches and contributes empirical insights to the fields of AI-assisted decision-making and human-AI collaboration. We encourage researchers to venture beyond conventional XAI-based support and to investigate a broader range of novel AI-assisted decision-making paradigms.

\textbf{Adapting AI Roles to Suit Varying Situations}. Our experiments indicate that a high-performing AI is most effective in the Recommender role. Conversely, when AI performance is low, the Analyzer role enhances task performance, fosters more appropriate reliance, and improves the user experience. Since AI's calibrated confidence can serve as an indicator of its accuracy, it is potentially effective to adaptively switch AI roles based on AI's confidence in a specific prediction. For instance, in scenarios where AI confidence is low (suggesting a lower probability of correctness), the AI could assume an analytical role, providing counsel rather than direct recommendation.

\textbf{Choosing AI Roles Based on Specific Objectives}. For instance, in high-risk decision-making scenarios where reducing over-reliance on AI is crucial, assigning AI the role of Devil's Advocate might be beneficial, even when AI performance is high. Conversely, to mitigate under-reliance on AI, the Recommender role may be more effective.

\textbf{Limitations and Future Work}. Our study's main limitation is its small participant sample. In the next step, we plan to expand the number of participants. Besides, in this paper, we only explored the interaction effects between AI roles and AI performance. Future research will explore the effects of the AI role with two critical factors: task difficulty and human-AI complementarity. Task difficulty affects both independent human performance and their response to various types of assistance, as well as their willingness to accept advice. Besides, the effectiveness of AI roles like Recommender, Analyzer, and Devil's Advocate may vary based on the degree of human-AI complementarity. Additionally, we plan to investigate how personal characteristics \cite{schwenk1982effects, bonaccio2006advice}, such as the need for cognition, ambiguity tolerance, self-esteem, and agreeableness, influence the effectiveness of different AI roles. This investigation will help us understand if different individuals are more suited to be accompanied by specific AI assistant roles. Moreover, in the next round of study, we will collect participants' open-ended feedback to qualitatively analyze their nuanced perceptions of different AI roles and gain a deep understanding of the root causes of the effects.


\section{Conclusion}

This position paper empirically investigates the impact of AI assuming varied roles in AI-assisted decision-making and its effects on human decision processes. Our initial experiments reveal that different AI roles exhibit distinct advantages and limitations, which are further influenced by the AI's performance. We aim for our findings to invigorate ongoing debates within the AI-assisted decision-making community regarding the optimal forms of AI assistance. Additionally, we hope to enrich discussions in the human-centered XAI community about the appropriate and human-compatible ways for AI to communicate explanations to users.

\balance

\bibliographystyle{ACM-Reference-Format}
\bibliography{main}


\end{document}